\documentstyle[twoside,fleqn,espcrc2]{article}

\newcommand{\AmS}{{\protect\the\textfont2
  A\kern-.1667em\lower.5ex\hbox{M}\kern-.125emS}}
\title{Quasi particles in hot QCD}

\author{P. Giovannangeli
        \thanks{Talk given by C. P. Korthals Altes}
        and
       C. P. Korthals Altes \address{Centre Physique Th\'eorique au CNRS, Campus de Luminy, \\ Case 907, 13288 Marseille Cedex, France }}

\begin{document}

\begin{abstract}
We show at very high temperature  how the behaviour of the spatial 
't Hooft loop in the QCD plasma is simply related to the chromo electric 
flux of  the gluons. This simple picture is vindicated by
a systematic quasi classical approach. The spatial Wilson loop 's behaviour is
computed by a similar nearly free plasma of magnetic quasiparticles.
This model predicts unambiguously ratios of multiply charged Wilson loops.
Recent simulations confirm these predictions accurately.

\end{abstract}

\maketitle

\section{Introduction}
Traditionally plasmas are a  state of matter in which electrically
charged constituents electrons and ions move - to a good approximation
- freely. What we know about the QCD plasma points in the same direction:
at extremely high temperatures quarks and gluons move freely.
Landau coined the word ``quasi-particles'' for nearly free collective
excitations and we shall adhere to this terminology for  quarks and gluons.

In this talk we report on the existence of a  third component in a QCD plasma:
magnetically charged quasiparticles. They explain the rise with temperature
in the
surface tension of the spatial Wilson loop.  
They unambiguously predict the ratios of surface tensions of multiply 
charged Wilson loops. And these have been verified in numerical simulations  
to about one percent accuracy.

The next sections explain the idea. In the last section we propose additional 
tests.

\section{The QCD plasma: some basic facts}

What the lattice has taught us about the plasma can be summarized for our purposes in the behaviour of two quantities: the Wilson loop and the 't Hooft loop.

As we will be interested in the flux emanating from quasi particles we concentrate on the space like loops. In the plasma phase we have screening.  This does  exclude area law behaviour for time like, but not for spacelikeloops. 
      
The space-like Wilson loop measures the magnetic flux. For an SU(N) group one can distinguish multiply charged loops: $W_k(L)$. k is the number of quarks modulo $N$ needed to build the
representation used in the loop. For the arealaw only k mod N matters.
This follows basically because of the relationship between Wilson and 't Hooft loops that loop each other once \cite{thooft}:
\begin{equation}
V_{k'}(L')W_k(L)V_{k'}^{\dagger}(L')=\exp{ikk'{2\pi\over N}}W_k(L)\,.
\label{eq:comrel}
\end{equation}
The integer $k'$ is the charge of the 't Hooft loop, i.e. the discontinuity
across the surface spanned by $L$ is $\exp{ik'{2\pi\over N}}$.
An eventual area law will only depend on the centergroup appearing in this
commutation relation.

Now the facts. The tension $\rho$ of the Wilson loop in SU(3) (and earlier SU(2)) has been measured to be flat
in the cold phase~\cite{boyd}, and to start growing at about $T=2T_c$ like one expects in the three dimensional reduced magnetic QCD, where only one scale, the magnetic scale $g^2T$ survives
\begin{equation}
\rho(T)\sim(g^2T)^2 \,.
\end{equation}
The magnetic mass, defined as the screening mass in the force law between
two heavy magnetic monopoles~\cite{rebbi} is non-zero at $T=0$~\cite{thooft}
and stays constant~\cite{rebbi} till about $T=2T_c$ and then follows again a 
behaviour as expected in the 3d theory mentioned above.

The 't Hooft loop has been measured only recently in SU(2), near $T=T_c$~\cite{deforcrand}. It shows an area law $\exp{-\sigma(T)A}$, as expected from semiclassical considerations~\cite{kovner} at high T. The tension obtains from  tunneling from a minimum where the Wilson line $P={1\over N}Tr{\cal{P}}\exp{i\int d\tau A_0}$ has the trivial value $P=1$ to a Z(N) symmetry conjugated minimum where the value is $P=\exp{ik{2\pi\over N}}$. We get a tension $\sigma_k(T)\sim k(N-k)\sigma_1$ for the multiply charged loop, up and including two loop order. For the singly charged 't Hooft loop one has~\cite{kovner}:
%\begin{equation}
\begin{eqnarray}
\sigma_1={4\pi^2T^2(N-1)\over{3\sqrt{3g^2(T)N}}}
       (1-(11.9905){g^2(T)N\over{(4\pi)^2}}
       \\
       +O(g^4)) \,.  \nonumber
\label{eq:defor}
\end{eqnarray}
%\end{equation}
No dynamical quarks are involved in these simulations. Their implementation is 
quite difficult, and notably area behaviour persists as a meta stable state
even when   perimeter behaviour is the stable state as in the case of time like Wilson loops. 

\section{Explanation of area law behaviour of 't Hooft loop with gluonic quasi particles}

The 't Hooft loop  measures the colour electric flux of physical states:
\begin{equation}
V_1(L')=\exp{i{4\pi\over N}\int_{S(L')} Tr\vec E Y_1 d\vec S}
\label{eq:thloop}
\end{equation}
where $Y_1$ is the familiar hypercharge, an NxN traceless diagonal matrix with the first N-1 entries
equal one and the last entry equal to $-N+1$. This operator (\ref{eq:thloop}) is in the physical subspace of Hilbert space  a correct representation of the usal definition
as a gauge transformation with a discontinuity $exp{i{2\pi\over N}}$ across the area $S(L')$.

 For the k times charged 't Hooft loop we have
\begin{equation}
V_k(L')=\exp{i{4\pi\over N}\int_{S(L')} Tr\vec E Y_k d\vec S}
\label{eq:thkloop}
\end{equation}
with $Y_{k}$ diagonal and traceless with $N-k$ times k and k times $N-k$
on the diagonal. Note that we do not take $kY_1$ because that would have given $k\sigma_1$ and from the semiclassical calculation we know~\cite{giovanna} the correct answer
is $k(N-k)\sigma_1$. The matrix $Y_{k}$ is closer by from the tunneling point of view. 

The flux is provided by the gluonic quasi particles in the adjoint representation. They have hypercharge $0,\pm N$, as follows from the adjoint representation of $Y_1$. The hypercharge $Y_{k}$ in the adjoint rep. has $k(N-k)$ entries with $N$ and an equal number with $-N$ the rest being zero.

Now the contribution to the 't Hooft loop due to one gluon is $1$ if it has no charge and $-1$ if it has charge. This is readily seen from the commutation
relation with the Wilson loop in the adjoint representation. We simplify by assuming the gluon contributes only within a thin slab of thickness $l_D$, the Debye screening length. 

To calculate the average one has to know the thermal distribution $P(l)$ of the gluons inside the slab.  Due to SU(N) global symmetry $P(l)$ is the same for all species of gluons. The average of the loop becomes :
\begin{equation}
<V_k>=\exp(-2k(N-k)\gamma w(T)) \,.
\end{equation}
$w(T)=c \bar l$ is the width of the distribution and since it thermodynamic the
width is proportional to the mean number $\bar l$ of gluons in the slab.
And the latter equals $l_DL^2n(T)$, n the density of one gluonic species.
The constant $\gamma$ is a geometric factor correcting for the simplifying assumption that a gluon only contributes flux if within the Debye length.
The factor $2k(N-k)$ follows from counting the gluons with charge $\pm N$ in the adjoint
representation of $Y_k$.

So we found the area law, and how it depends on the charge $k$ of the loop.
The constant $c$ in the thermodynamic distribution and the geometric factor $\gamma$ do clearly 
not depend on $k$. $k$ and $k+N$ give the same operator. This periodicity,
together with charge conjugation invariance, $k\to -k$, explains the k-dependence.

\section{Magnetic quasi particles as  an explanation for the rise in the Wilson loop}

The Wilson loop does not admit a quasi classical calculation.
But the rise in the tension in the hot phase does clearly indicate there is additional magnetic activity. Taking heart from our previous calculation we are going to assume the simplest possibility:
in the plasma phase there are screened (screening length $l_M$) magnetic charges with a density $n_M$ such that $l_M^3n_M<<1$. If so these charges are behaving like a free gas. We must decide on the internal degrees of freedom of the charges.
     At very high T the monopole gas reduces to a gas of lumps in the local action density of d=3 SU(N) gauge theory. The average action density is known
to scale like $N^2-1$. So the simplest guess is that the monopoles are in the
adjoint representation of a  magnetic SU(N) group.  This is consistent with the
results of Goddard et al. \cite{goddard} for classical non-abelian
monopoles. Of course our magnetic quasi particles are far from classical,
they are collective excitations of the plasma. 

This magnetic group needs only to be  global to  have the same distribution law for all $N^2-1$ magnetic quasiparticle species.
 
Arguing as above for the gluon gas, we find now that the Wilson tension
in a loop built from a representation with k quarks will behave like:
\begin{equation}
\rho_k=dk(N-k)l_M n_M
\label{eq:final}
\end{equation}
and the constant drops out in the ratio 
\begin{equation}
{\rho_k\over{\rho_1}}={k(N-k)\over{(N-1)}}
\label{eq:kratio}
\end{equation} 
From the data by Boyd et al.~\cite{boyd} we know that
 $l_Mn_M\sim g^4T^2$
or $n_M\sim l_M^{-1}T^2=g^6T^3$. Thus the magnetic quasi particles do not show up in the 
Stefan Boltzmann limit. Their density being $~({T\over{\log T}})^3$, they must have residual interactions, that we neglected in obtaining eq.(\ref{eq:final}).

Lucini and Teper~\cite{teper} checked the ratios $(\ref{eq:kratio})$ for Wilson loops and found very good 
agreement for SU(4) and SU(6), within about a percent. 

\section{Prospects}

Although the results for the ratios are very encouraging (the SU(5) case is
under study right now~\cite{giacomo}), we would like to get the constants 
mentioned above, the ratio of screening length to interparticle distance
 and an idea about the interactions.

An obvious task is to understand eventual d=3 local action lumps. Some ten years
ago~\cite{duncan} Duncan and Mawhinney found, after cooling, lumpy behaviour
of the action density in the SU(2) case. It is imperative to  understand
and quantify this behaviour, either by lattice means or by understanding the transition
from the semi classically accessible Higgs phase to the unbroken SU(2)
sytem by decoupling the Higgs.  The numerical results~\cite{duncan} seem to indicate that 
the 't Hooft Polyakov monopoles delocalize in terms of the lattice length, when decoupling the Higgs. An analysis of the  continuum limit is in order.

In the presence of dynamical fermions, spatial Wilson loops should behave
qualitatively the same in the hot phase. This is due to the quarks being Dirac compatible with the magnetic quasi particles. Since it is the 
magnetic component that is responsible for the area law we expect no
qualitative change. In the cold phase the tension disappears, whereas the't Hooft loop acquires a tension~\cite{kovner}.

Presently we are calculating the $O(g^4)$ term in the 't Hooft loop, eq.(\ref{eq:defor}). Indications are that it again behaves as $k(N-k)$. It may be
that this ratio is constant in the deconfined phase.   

The early onset (at $T\sim 2T_c$) of the rise of the Wilson loop tension~\cite{boyd} suggests the new component may  improve the behaviour of the pressure resulting from only the gluonic component.

\end{document}